\documentclass[11pt,twoside]{article}
\usepackage{asp2010}

\resetcounters

\begin{document}

\title{Why Galaxies Care about AGB Stars. Modelling Galaxies.}
\author{C. Maraston
\affil{Institute of Cosmology and Gravitation, University of 
Portsmouth, U.\,K.}}

\begin{abstract}

The Thermally-Pulsating Asymptotic Giant Branch (TP-AGB) phase of 
stellar evolution has received attention only recently in galaxy 
evolution, but is now an important player in our understanding of 
how galaxies form and evolve. Because it is a short but very luminous 
phase, bright in the near-IR where dust effects are small, the 
TP-AGB phase is a powerful tracer of intermediate-age stars in 
galaxies up to high redshift. The spectral signature of TP-AGB 
stars as defined by population synthesis models has been detected 
by the {\sl Spitzer Space Telescope} in high-redshift galaxies, whose 
spectra show an amazing similarity to spectra of local stellar 
populations. Even accounting for the high uncertainty affecting 
the theoretical modelling of this phase, stellar population models 
including the TP-AGB have leveraged a better determination of galaxy 
ages and hence stellar masses, fundamental quantities for 
studying galaxy formation and evolution. They have also improved 
the results of semi-analytic models, which can better reproduce 
colours and the $K$-band luminosity function of high-$z$ galaxies. 	
\end{abstract}

\section{Background: Stellar Population Models and Galaxy Evolution}

Stellar population models are the tool to perform galaxy evolution 
studies, as they predict the integrated spectrophotometric 
properties of arbitrary populations of stars as a function of 
parameters such as age, chemical composition, star formation 
history, initial mass function, stellar mass, etc.  Their 
usage is twofold: they allow the derivation of galaxy properties 
from data on observed galaxies, and they are the ingredients for 
calculating the predicted spectra of synthetic galaxies from 
galaxy formation models \citep[see e.g.][]{bau06}. I will review 
results on both aspects, focusing on the role played by the 
Asymptotic Giant Branch (AGB) phase of stellar evolution -- the 
protagonist of this meeting -- on our understanding of galaxy 
evolution.

\subsection{Model Basics}

Stellar population models are calculated assuming (1) stellar 
evolution models which provide the energetics at given stellar mass; 
(2) stellar spectra for distributing the energetics at the various 
wavelengths; and (3) a numerical algorithm to calculate the 
integrated spectral energy distribution (SED). This approach -- 
which is based on stellar evolution -- is referred to as 
Evolutionary Population Synthesis (EPS), the foundations of which 
are due to the pioneering work by \citet{tin72} and \citet{ren81}.
\begin{figure}[!ht]
\begin{center}
   \includegraphics[bb=50 150 750 580,width=0.9\textwidth]{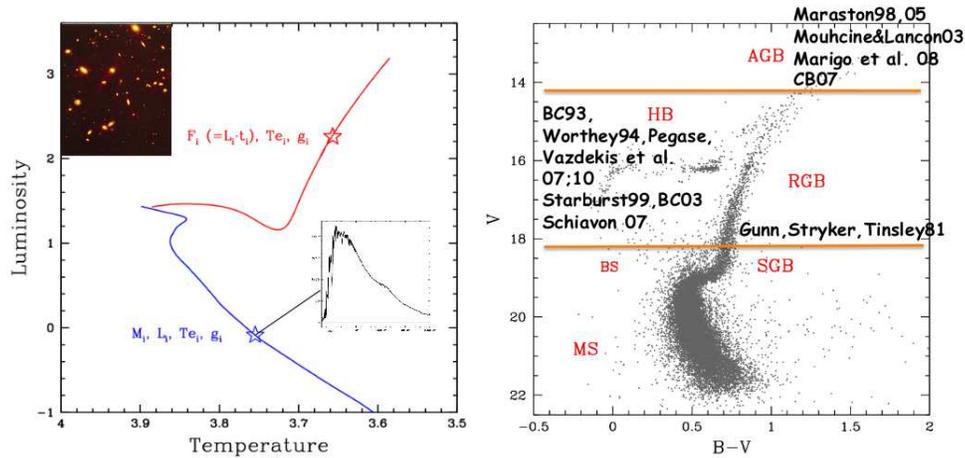}
   \caption{EPS model calculation and model evolution over years. 
{\it Left}:  Theoretical HR diagram for a coeval 
population, with blue and red lines indicating stars on the Main 
Sequence (MS) and post-Main Sequence. The inserts show a galaxy
cluster at redshift 0.6, and the spectrum of a MS star. 
The integrated spectrum is calculated by summing 
up the contribution by all mass bins, following different techniques 
(see text). 
{\it Right}:  The evolution of EPS models with respect to 
the inclusion of stellar evolutionary phases. The TP-AGB has been 
added last to population synthesis.}
  \label{eps}
  \end{center}
\end{figure}
%

Figure \ref{eps} illustrates the concept of EPS models and their 
development over the last three decades. The left-hand plot shows 
the theoretical HR diagram for a coeval population of stars with 
different initial masses, up to the tip of the Red Giant Branch 
(RGB). The spectrum of the population is obtained by integrating 
the luminosity contribution from stars of different masses convolved 
with the assumed initial mass function (IMF). 

The integral can be performed using mass as the evolutionary variable, 
as in the so-called isochrone synthesis \citep{brucha93}.
An alternative approach is based on Renzini's {\it fuel 
consumption theorem} \citep{ren81}, as in the models by \citet{buz89} 
and \citet[][hereafter M98, M05]{cm98,cm05}, and also by 
\citet{margir07} for constraining the contribution of the TP-AGB 
in their stellar models. In the Maraston models, the integration 
variable adopted in post-MS is the {\it fuel}, i.e. the product of 
luminosity and lifetime (Fig.~1, {\it left}). This 
approach turned out to be particularly useful for inserting a 
prescription for the TP-AGB in EPS models.
The right-hand panel provides a historical overview over the 
evolution of these models, which could be viewed as the evolution 
in the number of major stellar phases that could be included in the 
synthesis. While at the end of the 1970's models could cover just the 
base of the RGB \citep{gunetal81}, during the 1990's the availability 
of large grids of isochrones from the Padova group \citep{beretal94} 
allowed the inclusion of all stellar phases up to the end of the 
Early-Asymptotic Giant Branch 
\citep[E-AGB;][]{wor94,brucha93,vazetal96,vazetal10,pegase,sb99,sch07} 
or shortly after as in the models by Bruzual \& Charlot (2003, 
hereafter BC03). 

The TP-AGB was included in the Maraston models in a semi-empirical 
fashion, as described in the next Section. \citet{maretal08} 
extend the Padova isochrones with their models for the TP-AGB, after 
calibrating them with data as in the Maraston models. 
\citet{moulan03} present an implementation of TP-AGB based on their 
own stellar models. The reader is referred to the review by 
Lan{\c c}on (this volume) for further details on models and comparison 
to star clusters.

\subsection{TP-AGB in Integrated Models: a Semi-empirical Approach}

The theoretical modelling of the TP-AGB presents several well-known 
but challenging aspects stemming from the pulsational regime, the 
double-shell burning, and especially the strong mass loss affecting 
this phase.  Also complicated 
is the spectral modelling, especially during the carbon-rich phase. 
These issues have traditionally hampered the calculation of isochrones 
that could easily be implemented in population synthesis. 
%
\begin{figure}[!ht]
\begin{center}
   \includegraphics[bb=50 60 750 590,width=0.9\textwidth]{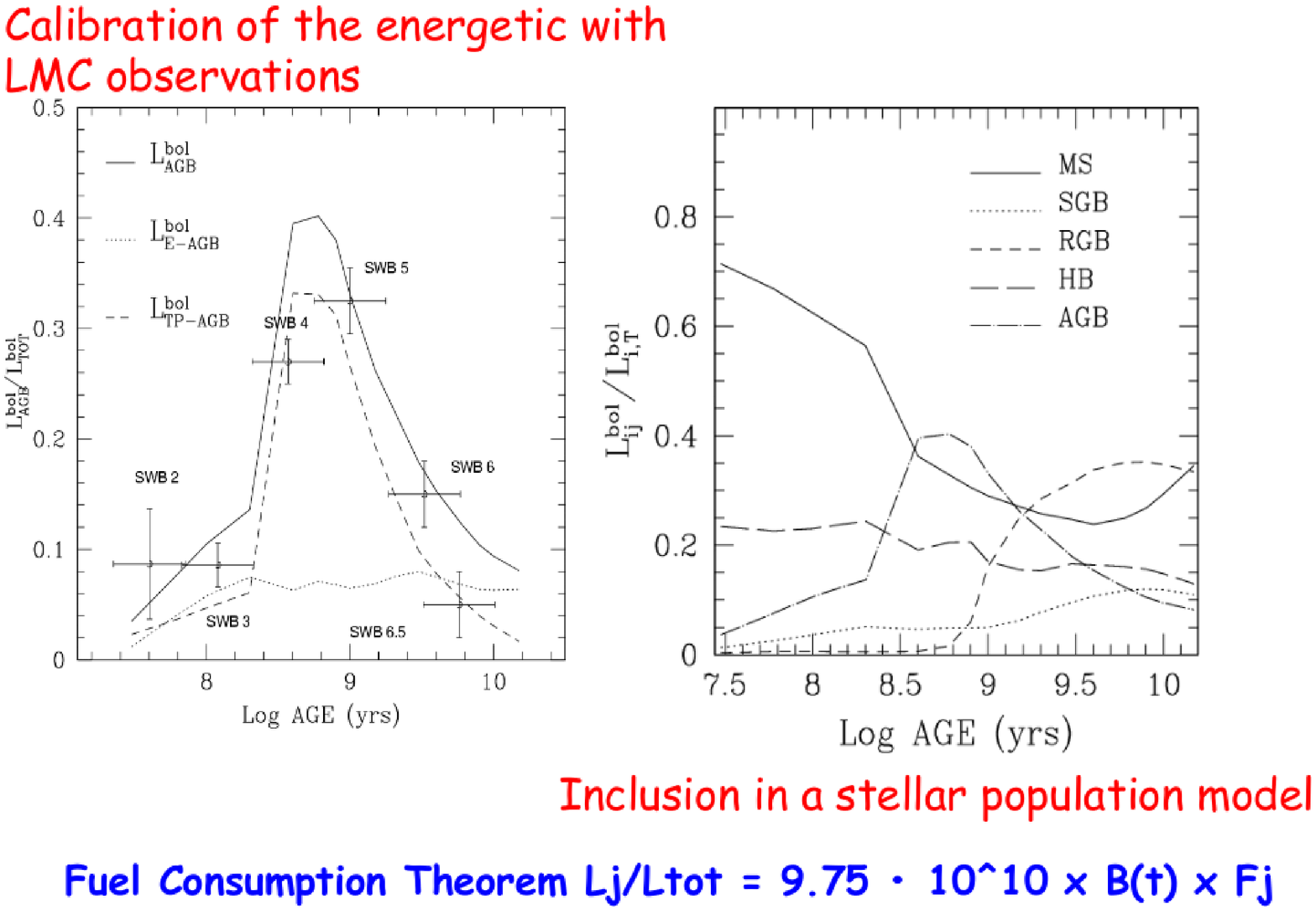}
   \caption{Modelling the TP-AGB with the fuel consumption theorem}
  \label{empm98}
  \end{center}
\end{figure}
In order to overcome this situation and test the effect these 
luminous stars have on the integrated properties of galaxies, 
M98 attempted a semi-empirical approach. 
The theoretical energetics in the TP-AGB phase for stellar models of 
several initial masses were calibrated with the observed bolometric 
contribution by the TP-AGB in Magellanic Cloud (MC) globular 
clusters (GCs) from \citet{froetal90}. Key to this calibration is 
the fuel consumption theorem \citep[FCT;][see Eq. in Figure 3]{ren81}, 
which expresses the bolometric luminosity of a post-MS phase $J$~in 
terms of the {\it fuel} (i.e. equivalent masses of H and/or Helium) 
at disposal for core/shell burning to a star of mass equal to the 
turnoff mass $M_{\rm TO}$~at the given population age. Note that 
because the fuel is calibrated with observations, it automatically 
includes the (otherwise unknown) effect from mass-loss. 
The left-hand plot of Figure 2 shows the calibration of the bolometric 
contribution of the AGB phase to the total as a function of the age 
of the population, split into E-AGB and TP-AGB. This calibration 
shows a peak contribution by the TP-AGB phase in populations with 
ages $\sim\,0.5-1$~Gyr, by up to 40\% of the bolometric contribution. 
It is important to stress that stochastic fluctuations in the number 
of bright AGB stars per cluster, which are due to the shortness of 
the TP-AGB phase ($\sim\,10^6\rm yr$), were minimised by averaging 
the TP-AGB contributions in GCs of similar age (see M98 for 
details). Vertical error bars show the size of such stochastic 
fluctuations. Horizontal error bars represent the uncertainty in 
GC ages, which stem from the fact that tracks with and without 
overshooting fit the same turnoff with ages that differ by a few
hundred million years \citep{giretal95,feretal95,feretal04}.
Note also that the MC GCs are the most useful calibrators in this 
context as their metallicities and ages -- unlike more distant star 
clusters or galaxies -- can be derived independently of stellar 
population models.
\begin{figure}[!ht]
\begin{center}
   \includegraphics[bb=50 100 750 600,width=0.9\textwidth]{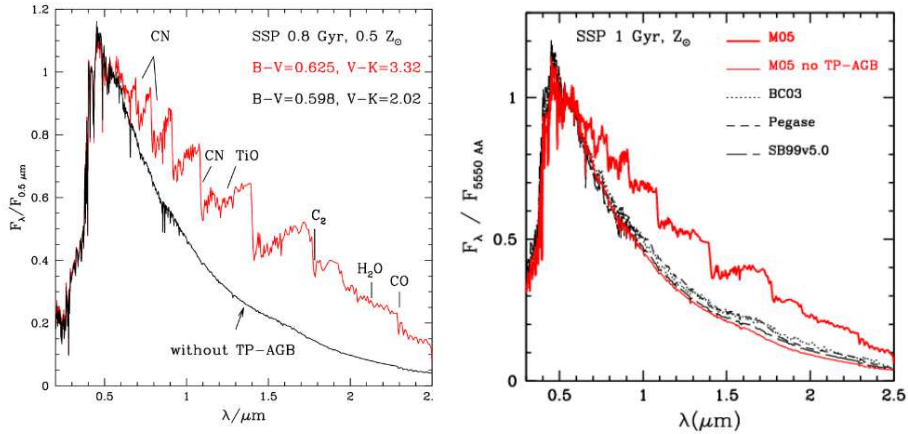}
   \caption{Effect of TP-AGB on the synthetic SED of population models}
  \label{m05sed}
  \end{center}
\end{figure}

M05 -- following the availability of empirical spectra of C-rich 
and O-rich TP-AGB stars from \citet{lanwoo00} -- extended 
the modelling of M98 by calculating the full SED of population 
models of various ages and metallicities, with the aim of studying 
high-redshift galaxies. Figure~\ref{m05sed} (left) 
shows the effect of the TP-AGB on the integrated SED. The 
addition of luminous red stars increases the flux in the near-IR 
compared to a model including only the E-AGB, and the spectral 
absorptions typical of TP-AGB stars become apparent. The right-hand 
plot compares model SEDs from different authors, not including the 
TP-AGB or adopting different prescriptions (see M05 for details and 
references). This plot helps understanding discrepant results that 
are obtained on galaxies when different models, together with data 
extending to the near-IR rest-frame, are used to derive galaxy 
properties. 

\subsubsection{Caveats}

There are a number of caveats intrinsic to the semi-empirical 
modelling, namely: the onset age, the energetics, stochastic
fluctuations, and metallicity effects. 

First, the age at which the TP-AGB is held to be relevant is calibrated 
with the ages of MC GCs, which depend on the tracks adopted to derive 
them and also on the fitting method. 
In the Maraston models the relevant age range is between 0.3 and 2 Gyr 
(Fig.~2), which stems from turnoff ages derived via classical 
(non overshooting) tracks. Recent work has confirmed this scale 
\citep{mucetal06}. Tracks with overshooting make the peak age older 
by some hundred million years \citep{giretal95,maretal08}. 
However, there is not yet universal agreement on the actual size 
of overshooting. Moreover, MC age fitting that includes post-MS stars 
and not only the turnoff, pushes the ages of some MC GCs to even older 
values \citep{keretal07}. This older age scale was used by 
\citet{pesetal08} and \citet{cg10} to argue that the Maraston models 
are mis-calibrated, but this -- more than a conclusion -- is a 
tautology as the calibration depends on the adopted age scale. 

Stochastic fluctuations in the number of bright AGB stars affect the 
observed colours and magnitudes (see Lan{\c co}n, this volume). 
M98 used a representative average cluster for each age bin (Fig.~2, 
details in M98) so that there is probably little doubt that the 
bolometric contribution of the TP-AGB can be up to 40\,\%, whereas 
the maximum age and the shape of the phase transition depends on the 
adopted age scale and on the age binning.

A further caveat regards the metallicity dependence of the energetics 
and stellar spectra. The Maraston models assume the theoretical 
scaling proposed by \citet{renvol81} according to which a metal-rich 
population burns more TP-AGB fuel in form of oxygen-rich stars than 
in carbon stars. This trend of spectral type with metallicity is 
confirmed by modern models \citep[e.g.][]{maretal96}, though the 
quantitative scaling may be different. 
On the other hand, the same empirical spectra are used, which involves 
the assumption that the main spectral features of the two types do not 
depend on the initial chemical composition. \citet{lyuetal10} 
cast some doubt on this assumption, as they find that the CO 
absorption measured in two MC GCs with ages around 1 Gyr is lower 
than predicted by the M05 models (while the integrated colours of 
the same objects perfectly agree). Their suggestion is that -- due 
to the lower metallicity of the Magellanic Clouds with respect to 
our Galaxy and the stars observed by Lan{\c c}on \& Wood  
-- the spectral absorptions cannot be adequately modelled. More 
data will be interesting, as the result is based on only two objects 
with the lowest S/N. 

\section{Why Real Galaxies Care about AGB Stars}

The AGB phase is bright and well-populated in intermediate-age stellar 
populations for a short period of time. This makes it a powerful age 
indicator, and generally a useful tool to break the age/metallicity 
degeneracy, which plagues the analysis of galaxy spectra. The 
high-redshift Universe is the best target for several reasons: 
galaxies are younger and the age spread between different stellar 
generations is smaller than in local objects, both facts helping 
to enhance the signal of the AGB phase in the spectra. The M05 models 
were motivated by the launch of NASA's 
{\sl Spitzer Space Telescope}, which could sample the galaxy rest-frame 
near-IR spectrum -- which hosts the signature of red stars 
(cf.\ Fig.~3) -- up to high redshift \citep{cm05}. A large number 
of papers have been published on the topic. In the following we 
shall provide a brief snapshot of the importance of the TP-AGB in 
galaxies as a function of redshift. 

\subsection{AGB Stars: Common Inhabitants as a Function of Redshift}

Figure 4 illustrates the effect of the TP-AGB on the fitting of observed 
galaxy spectro\-photometry, for galaxies at various redshifts and for 
local MC GCs. 

\begin{figure}[t]
  \epsscale{1.}\plotone{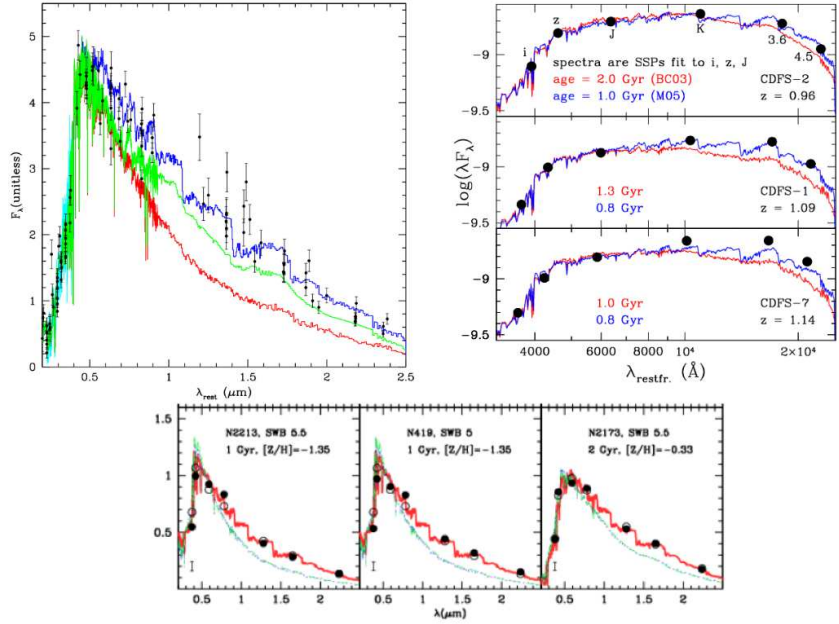}
  \caption{AGB stars are common inhabitants in high-redshift galaxies. 
{\it Upper left}:  Stacked spectrum of $z\sim2$~galaxies from 
\citet{cimetal08}, compared to models by M05 (blue), BC03 (red) 
and Charlot \& Bruzual (2007, {\it private communication}, green). 
{\it Upper right}:  
Data for field galaxies at redshift 1 fitted with M05 and BC03 models 
\citep[red and blue lines, from][]{vdwetal06b}. {\it Bottom panel}: 
Large Magellanic Cloud GCs SED fit from M05.}
  \label{cb07}
\end{figure}

The right-hand plot shows the SED fit for galaxy data at redshift 
$\sim1$ with M05 and BC03 models (blue and red lines, respectively). 
The SED fit is performed using just the three bands $i,z,J$, which 
sample the rest-frame spectrum up to $\sim8000$~\AA. Longer 
wavelength data are added onto the best-fit model, without using 
them to further constrain the fit. The M05 models with their 
prescriptions for the TP-AGB phase fit the rest-frame near-IR much 
better, the other model lacking near-IR rest-frame flux.
Note that the galaxy light-averaged ages obtained with the two 
models are different, namely the M05 models give younger ages.
Finally, note that if one fits all six bands simultaneously, 
good solutions would be found for the BC03 models as well, which 
probably would yield older ages/higher metallicities/higher dust 
content, in essence all model ingredients that help to give a 
higher flux in the near-IR. One cannot avoid noting, however, 
that the TP-AGB solution fits the whole spectrophotometry well. 
The left-hand panel shows a similar SED fit for a galaxy stack 
at $z\sim2$. Note the strong near-IR fluxes of these galaxies that 
are best fitted with models including a strong TP-AGB. The figure 
also includes a comparison with the models by Charlot \& Bruzual 
({\it in preparation}) which adopt the \citet{maretal08} isochrones.  
Finally note the SED fits of MC GCs from M05. The similarity between 
the spectra of small, local objects, and distant, massive galaxies 
with similar ages is suggestive of an overall similarity in the 
stellar evolution of AGB stars born locally and in the distant 
Universe. 

\begin{figure}[!ht]
\begin{center}
   \includegraphics[bb=60 90 750 580,width=0.9\textwidth]{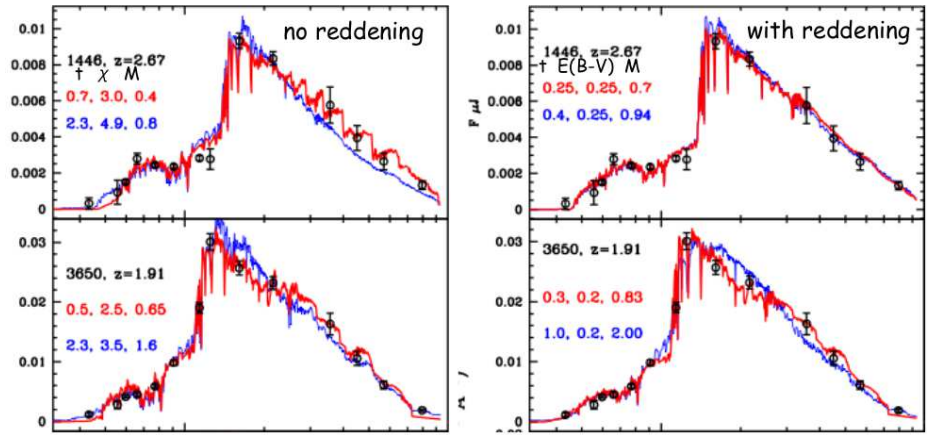}
   \caption{Effect of population models -- M05 in red and BC03 in blue -- on the 
physical properties derived for galaxies. {\it Left}: the
SEDs of two galaxies on the assumption of no reddening; 
{\it right}: the same galaxies fitted with additional reddening. 
Labels indicate age, $\chi^2$ and stellar mass on the left, and age, 
reddening $E(B-V)$ and stellar mass on the right. Note that these 
galaxies were pre-selected to be mostly passive and were found to be 
lacking dust re-emission; hence the fit with reddening is a formal, 
but probably not very realistic, solution. 
From \citet{cmetal06}}
  \label{mass}
  \end{center}
\end{figure}

We now illustrate why details of the TP-AGB modelling are critical 
for the derivation of galaxy properties. Figure~\ref{mass} shows 
the SED fit of galaxies at redshift $\sim 2$~for the M05 (red) and 
the BC03 (blue) models. In the left-hand panel reddening was not 
included in the fit. Galaxy ages derived with the M05 models are 
lower, because the TP-AGB makes the galaxy red at a younger age, 
whereas the BC03 models require older ages with a pronounced RGB 
for matching the high near-IR fluxes, even if the fit is not 
equally good. The older ages explain the higher masses. The 
addition of reddening dilutes, but does not remove, the differences. 
\citet{cmetal06} found that masses derived with M05 are on average 
$\sim$\,30 to 50\,\% lower than those derived with BC03, a figure that 
has been confirmed in several subsequent works. It is important to 
note that these galaxies have been found to have a negligible amount 
of dust. Hence, the solutions with reddening are statistically good, 
but perhaps not very realistic. Similar results can be found in the 
literature, leading to the following conclusions: models with 
TP-AGB give lower ages, lower stellar masses and lower dust content.

One may be tempted to conclude that a TP-AGB prescription as in M05 
provides good fits to galaxies over a wide redshift range. This 
conclusion has been challenged by \citet{krietal10}, who -- based 
on their SED fit of colour-defined post-starburst galaxies with 
photometric redshifts over a broad redshift range (1 to 2.5) -- 
conclude that M05 and \citet{maretal08} overestimate the TP-AGB 
contribution while the BC03 models work better. It is not easy to 
reconcile this result with others - for example \citet{raietal11}
just conclude the opposite, namely that M05 and Charlot \& Bruzual 2007 {\it private communication}
fit better than BC03 (see also the next Section). 

Finally, worth mentioning in the high-redshift context is the result 
by \citet{keletal10} who -- extending the M05 models longward to the 
$K$-band -- find that the contribution of AGB stars to dust re-emission 
is significant and affects the determination of galaxy star-formation 
rates from 24 $\mu$m data, lowering the high values sometimes 
derived for high-$z$ galaxies, with obvious impact on our understanding 
of galaxy evolution. 
 
\subsection{Local Galaxies in Integrated Light}

Several types of galaxies in the local Universe are known to host 
significant numbers of AGB stars, and this conference has an entire 
session devoted to resolved stellar populations in nearby systems. 
Here I will focus on results for the local Universe that are obtained 
in integrated light. For galaxies with star formation -- or, more 
generally, with stellar populations with ages around 1 Gyr -- the 
modelling of the TP-AGB should be relevant.

For spiral galaxies, \citet{macetal10} find that the star formation 
histories derived from optical spectroscopy with the BC03 models 
give predicted near-IR colours that are too blue compared with the 
observed ones. The correct colours, matching the observations, are 
obtained with the M05 models because of their brighter TP-AGB.
Eminian et al. (2008) analyse the near-IR colours of a sample of 
$\sim\,6000$ galaxies from SDSS and find that galaxies with higher 
star formation rates have bluer optical colours {\it and} redder 
near-IR colours, which are better explained with models including 
the TP-AGB phase such as M05 and CB07 rather than the BC03 ones. 
Both conclusions are at odds with the results by \citet{krietal10} 
mentioned above.
From the spectroscopic side, \citet{rifetal07,rifetal08} obtained 
the near-IR spectra of Seyfert/AGN-host galaxies and show that the 
spectral bumps in the near-IR can only be explained with the M05 
models because they include empirical carbon- and oxygen-rich  
TP-AGB stars. The amount of TP-AGB fuel in the M05 model does not 
appear to be excessive. Similarly, \citet{minetal11} measure 
C$_{2}$ and other spectral features in the near-IR spectra of the 
dwarf elliptical M\,32 and the post-starburst galaxy NGC 5102, finding 
that ages obtained with the M05 models are consistent with ages 
derived from the optical spectra in the literature. Moreover, they 
find that decreasing the TP-AGB fuel as suggested by \citet{krietal10} 
breaks this agreement (J. Miner, {\it private communication}). 
TP-AGB stars even affect galaxy dynamics. \citet{rotfis10} show 
that the measurement of dynamical masses for merger remnants from 
CO absorptions is affected by the choice of stellar population models 
and find encouraging results using the M05 models. In years to come 
we shall see the full development of near-IR spectroscopy which will 
greatly improve our knowledge on these issues.

\section{Why Semi-analytic Galaxies Care about AGB Stars}

\begin{figure}[!ht]
\begin{center}
  \includegraphics[width=105mm]{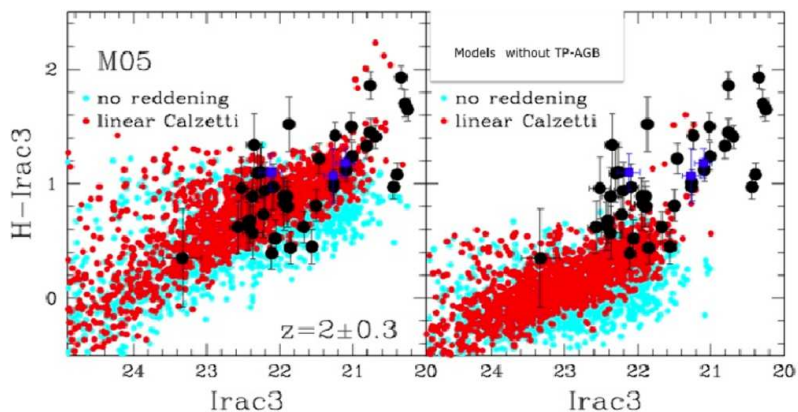}
  \caption{Observed-frame colour-magnitude diagram (corresponding to 
rest-frame $K$~vs. $V$--$K$) of redshift 2 galaxies from semi-analytic 
models \citep[small red and cyan points, from][]{tonetal10} compared 
to observations (large black circles).}
  \label{samcm}
  \end{center}
\end{figure}

Recent works have shown that the TP-AGB prescriptions in stellar 
population models adopted in galaxy formation models influence the 
meaning of the comparison with observations.
In \citet{tonetal09,tonetal10} we calculate semi-analitic models (SAM) 
with different input stellar population models and compare them to 
data for high-$z$~galaxies. Figure~\ref{samcm} shows that the 
observed-frame theoretical colour-magnitude diagram of semi-analytic 
galaxies at redshift 2 (small red and cyan points) is only able to 
match the observations (large black points) when the input stellar 
population model includes the TP-AGB phase (left-hand panel). This 
suggests that much of the well-known mismatch between high-$z$~data 
and traditional semi-analytic models (SAMs; right-hand panel) may be 
due to a light deficit at given mass, rather than to a mass deficit. 

\begin{figure}[!ht]
\begin{center}
   \includegraphics[bb=60 300 750 900,width=0.9\textwidth]{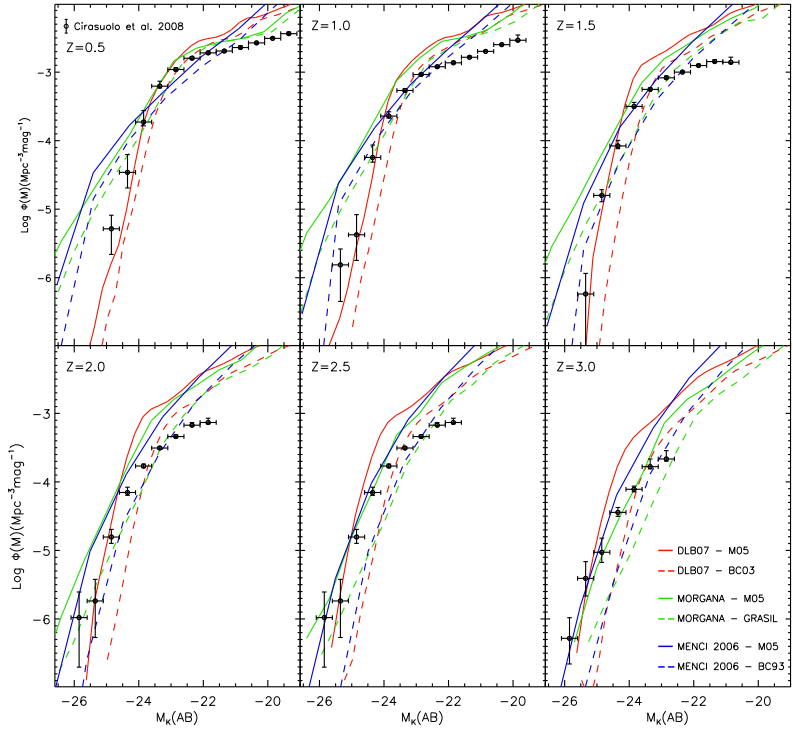}
\caption{Rest-frame $K$-band luminosity function of galaxies as a 
function of redshift \citep[filled points with errorbars,][]{ciretal10} 
compared with the predictions of several semi-analytic models as a 
function of the TP-AGB modelling in the input stellar population 
model. Dashed lines refer to stellar population models with little 
TP-AGB, solid lines to M05 models.}
  \label{samklf}
  \end{center}
\end{figure}

\citet{henetal11} use several SAMs equipped with models with and 
without the TP-AGB phase to understand the discrepancy with the 
observed near-IR high-$z$ galaxy luminosity function. 
Figure~\ref{samklf} shows that SAM models with the M05 models match 
the rest-frame near-IR luminosity function at high-redshift, where 
the discrepancy was pointed out. Finally, \citet{fonmon10} conclude 
that the observed colours of Extremely Red Objects can be better 
matched by SAM models including a bright TP-AGB.

\section{Concluding Remarks}

The recent inclusion of the TP-AGB in stellar population models has had 
a significant impact on our understanding of real galaxies as well as 
of synthetic galaxies from galaxy formation models. Galaxies do care 
about AGB stars! While the exact energetics for the TP-AGB may still 
be debatable, there is no doubt that several spectral features as well 
as the spectral continuum in the near-IR can be better  
explained with models including a substantial contribution 
(up to 40\%\ of the bolometric contribution) from the TP-AGB. Once more 
it is clear how our understanding of the galaxy formation process in a 
cosmological sense depends on our understanding of stellar evolution. 

\bibliography{b_maraston.bib} 

\begin{thebibliography}{}
\expandafter\ifx\csname natexlab\endcsname\relax\def\natexlab#1{#1}\fi
\expandafter\ifx\csname url\endcsname\relax
  \def\url#1{\texttt{#1}}\fi
\expandafter\ifx\csname urlprefix\endcsname\relax\def\urlprefix{URL }\fi
\providecommand{\eprint}[2][]{\url{#2}}

\bibitem[{{Baugh}(2006)}]{bau06}
{Baugh}, C.~M. 2006, Reports on Progress in Physics, 69, 3101

\bibitem[{{Bertelli} et~al.(1994){Bertelli}, {Bressan}, {Chiosi}, {Fagotto}, \&
  {Nasi}}]{beretal94}
{Bertelli}, G., {Bressan}, A., {Chiosi}, C., {Fagotto}, F., \& {Nasi}, E. 1994,
  \aaps, 106, 275

\bibitem[{{Bruzual} \& {Charlot}(1993)}]{brucha93}
{Bruzual}, G., \& {Charlot}, S. 1993, \apj, 405, 538

\bibitem[{{Buzzoni}(1989)}]{buz89}
{Buzzoni}, A. 1989, \apjs, 71, 817

\bibitem[{{Cimatti} et~al.(2008){Cimatti}, {Cassata}, {Pozzetti}
  et~al.}]{cimetal08}
{Cimatti}, A., {Cassata}, P., {Pozzetti}, L., et~al. 2008, \aap, 482, 21

\bibitem[{{Cirasuolo} et~al.(2010){Cirasuolo}, {McLure}, {Dunlop}
  et~al.}]{ciretal10}
{Cirasuolo}, M., {McLure}, R.~J., {Dunlop}, J.~S., et~al. 2010, \mnras, 401,
  1166

\bibitem[{{Conroy} \& {Gunn}(2010)}]{cg10}
{Conroy}, C., \& {Gunn}, J.~E. 2010, \apj, 712, 833

\bibitem[{{Ferraro} et~al.(1995){Ferraro}, {Fusi Pecci}, {Testa}
  et~al.}]{feretal95}
{Ferraro}, F.~R., {Fusi Pecci}, F., {Testa}, V., et~al. 1995, \mnras, 272, 391

\bibitem[{{Ferraro} et~al.(2004){Ferraro}, {Origlia}, {Testa}, \&
  {Maraston}}]{feretal04}
{Ferraro}, F.~R., {Origlia}, L., {Testa}, V., \& {Maraston}, C. 2004, \apj,
  608, 772

\bibitem[{{Fioc} \& {Rocca-Volmerange}(1997)}]{pegase}
{Fioc}, M., \& {Rocca-Volmerange}, B. 1997, \aap, 326, 950

\bibitem[{{Fontanot} \& {Monaco}(2010)}]{fonmon10}
{Fontanot}, F., \& {Monaco}, P. 2010, \mnras, 405, 705

\bibitem[{{Frogel} et~al.(1990){Frogel}, {Mould}, \& {Blanco}}]{froetal90}
{Frogel}, J.~A., {Mould}, J., \& {Blanco}, V.~M. 1990, \apj, 352, 96

\bibitem[{{Girardi} et~al.(1995){Girardi}, {Chiosi}, {Bertelli}, \&
  {Bressan}}]{giretal95}
{Girardi}, L., {Chiosi}, C., {Bertelli}, G., \& {Bressan}, A. 1995, \aap, 298,
  87

\bibitem[{{Gunn} et~al.(1981){Gunn}, {Stryker}, \& {Tinsley}}]{gunetal81}
{Gunn}, J.~E., {Stryker}, L.~L., \& {Tinsley}, B.~M. 1981, \apj, 249, 48

\bibitem[{{Henriques} et~al.(2011){Henriques}, {Maraston}, {Monaco}
  et~al.}]{henetal11}
{Henriques}, B., {Maraston}, C., {Monaco}, P., et~al. 2011, ArXiv e-prints.
  \eprint{1009.1392}

\bibitem[{{Kelson} \& {Holden}(2010)}]{keletal10}
{Kelson}, D.~D., \& {Holden}, B.~P. 2010, \apjl, 713, L28

\bibitem[{{Kerber} et~al.(2007){Kerber}, {Santiago}, \& {Brocato}}]{keretal07}
{Kerber}, L.~O., {Santiago}, B.~X., \& {Brocato}, E. 2007, \aap, 462, 139

\bibitem[{{Kriek} et~al.(2010){Kriek}, {Labb\'e}, {Conroy} et~al.}]{krietal10}
{Kriek}, M., {Labb\'e}, I., {Conroy}, C., et~al. 2010, \apjl, 722, L64

\bibitem[{{Lan{\c c}on} \& {Wood}(2000)}]{lanwoo00}
{Lan{\c c}on}, A., \& {Wood}, P.~R. 2000, \aaps, 146, 217

\bibitem[{{Leitherer} et~al.(1999){Leitherer}, {Schaerer}, {Goldader}
  et~al.}]{sb99}
{Leitherer}, C., {Schaerer}, D., {Goldader}, J.~D., et~al. 1999, \apjs, 123, 3

\bibitem[{{Lyubenova} et~al.(2010){Lyubenova}, {Kuntschner}, {Rejkuba}
  et~al.}]{lyuetal10}
{Lyubenova}, M., {Kuntschner}, H., {Rejkuba}, M., et~al. 2010, \aap, 510, A19

\bibitem[{{MacArthur} et~al.(2010){MacArthur}, {McDonald}, {Courteau}, \&
  {Jes{\'u}s Gonz{\'a}lez}}]{macetal10}
{MacArthur}, L.~A., {McDonald}, M., {Courteau}, S., \& {Jes{\'u}s
  Gonz{\'a}lez}, J. 2010, \apj, 718, 768

\bibitem[{{Maraston}(1998)}]{cm98}
{Maraston}, C. 1998, \mnras, 300, 872

\bibitem[{{Maraston}(2005)}]{cm05}
--- 2005, \mnras, 362, 799

\bibitem[{{Maraston} et~al.(2006){Maraston}, {Daddi}, {Renzini}
  et~al.}]{cmetal06}
{Maraston}, C., {Daddi}, E., {Renzini}, A., et~al. 2006, \apj, 652, 85

\bibitem[{{Marigo} et~al.(1996){Marigo}, {Bressan}, \& {Chiosi}}]{maretal96}
{Marigo}, P., {Bressan}, A., \& {Chiosi}, C. 1996, \aap, 313, 545

\bibitem[{{Marigo} \& {Girardi}(2007)}]{margir07}
{Marigo}, P., \& {Girardi}, L. 2007, \aap, 469, 239

\bibitem[{{Marigo} et~al.(2008){Marigo}, {Girardi}, {Bressan}
  et~al.}]{maretal08}
{Marigo}, P., {Girardi}, L., {Bressan}, A., et~al. 2008, \aap, 482, 883

\bibitem[{{Miner} et~al.(2011){Miner}, {Rose}, \& {Cecil}}]{minetal11}
{Miner}, J., {Rose}, J.~A., \& {Cecil}, G. 2011, \apjl, 727, L15

\bibitem[{{Mouhcine} \& {Lan{\c c}on}(2003)}]{moulan03}
{Mouhcine}, M., \& {Lan{\c c}on}, A. 2003, \aap, 402, 425

\bibitem[{{Mucciarelli} et~al.(2006){Mucciarelli}, {Origlia}, {Ferraro},
  {Maraston}, \& {Testa}}]{mucetal06}
{Mucciarelli}, A., {Origlia}, L., {Ferraro}, F.~R., {Maraston}, C., \& {Testa},
  V. 2006, \apj, 646, 939

\bibitem[{{Pessev} et~al.(2008){Pessev}, {Goudfrooij}, {Puzia}, \&
  {Chandar}}]{pesetal08}
{Pessev}, P.~M., {Goudfrooij}, P., {Puzia}, T.~H., \& {Chandar}, R. 2008,
  \mnras, 385, 1535

\bibitem[{{Raichoor} et~al.(2011){Raichoor}, {Mei}, {Nakata}
  et~al.}]{raietal11}
{Raichoor}, A., {Mei}, S., {Nakata}, F., et~al. 2011, ArXiv e-prints.
  \eprint{1103.0259}

\bibitem[{{Renzini}(1981)}]{ren81}
{Renzini}, A. 1981, Annales de Physique, 6, 87

\bibitem[{{Renzini} \& {Voli}(1981)}]{renvol81}
{Renzini}, A., \& {Voli}, M. 1981, \aap, 94, 175

\bibitem[{{Riffel} et~al.(2007){Riffel}, {Pastoriza},
  {Rodr{\'{\i}}guez-Ardila}, \& {Maraston}}]{rifetal07}
{Riffel}, R., {Pastoriza}, M.~G., {Rodr{\'{\i}}guez-Ardila}, A., \& {Maraston},
  C. 2007, \apjl, 659, L103

\bibitem[{{Riffel} et~al.(2008){Riffel}, {Pastoriza},
  {Rodr{\'{\i}}guez-Ardila}, \& {Maraston}}]{rifetal08}
--- 2008, \mnras, 388, 803

\bibitem[{{Rothberg} \& {Fischer}(2010)}]{rotfis10}
{Rothberg}, B., \& {Fischer}, J. 2010, \apj, 712, 318

\bibitem[{{Schiavon}(2007)}]{sch07}
{Schiavon}, R.~P. 2007, \apjs, 171, 146

\bibitem[{{Tinsley}(1972)}]{tin72}
{Tinsley}, B.~M. 1972, \aap, 20, 383

\bibitem[{{Tonini} et~al.(2009){Tonini}, {Maraston}, {Devriendt}, {Thomas}, \&
  {Silk}}]{tonetal09}
{Tonini}, C., {Maraston}, C., {Devriendt}, J., {Thomas}, D., \& {Silk}, J.
  2009, \mnras, 396, L36

\bibitem[{{Tonini} et~al.(2010){Tonini}, {Maraston}, {Thomas}, {Devriendt}, \&
  {Silk}}]{tonetal10}
{Tonini}, C., {Maraston}, C., {Thomas}, D., {Devriendt}, J., \& {Silk}, J.
  2010, \mnras, 403, 1749

\bibitem[{{van der Wel} et~al.(2006){van der Wel}, {Franx}, {Wuyts}
  et~al.}]{vdwetal06b}
{van der Wel}, A., {Franx}, M., {Wuyts}, S., et~al. 2006, \apj, 652, 97

\bibitem[{{Vazdekis} et~al.(1996){Vazdekis}, {Casuso}, {Peletier}, \&
  {Beckman}}]{vazetal96}
{Vazdekis}, A., {Casuso}, E., {Peletier}, R.~F., \& {Beckman}, J.~E. 1996,
  \apjs, 106, 307

\bibitem[{{Vazdekis} et~al.(2010){Vazdekis}, {S\'anchez-Bl\'azquez},
  {Falc\'on-Barroso} et~al.}]{vazetal10}
{Vazdekis}, A., {S\'anchez-Bl\'azquez}, P., {Falc\'on-Barroso}, J., et~al.
  2010, \mnras, 404, 1639

\bibitem[{{Worthey}(1994)}]{wor94}
{Worthey}, G. 1994, \apjs, 95, 107

\end{thebibliography}

\end{document}